%% file: braga.tex
\documentclass[12pt]{article}
\usepackage{graphicx}

\newcommand\pubnumber{}
\newcommand\pubdate{}

\def\institute{Laborat\'orio de Instrumenta\c{c}\~ao e F\'isica Experimental de Part\'iculas\\
LIP Lisbon, Portugal}

\def\Title#1{\begin{center} {\Large #1 } \end{center}}
\def\Author#1{\begin{center}{ \sc #1} \end{center}}
\def\Address#1{\begin{center}{ \it #1} \end{center}}

\newcommand\pubblock{\rightline{\begin{tabular}{l} \pubnumber\\
         \pubdate  \end{tabular}}}
\newenvironment{Abstract}{\begin{quotation}  }{\end{quotation}}
\newenvironment{Presented}{\begin{quotation} \begin{center} 
             PRESENTED AT\end{center}\bigskip 
      \begin{center}\begin{large}}{\end{large}\end{center} \end{quotation}}
\def\Acknowledgements{\bigskip  \bigskip \begin{center} \begin{large}
             \bf ACKNOWLEDGEMENTS \end{large}\end{center}}

\input econfmacros.tex

\begin{document}
\begin{titlepage}
\pubblock

\vfill
\Title{Latest Results on Top Quark Properties: \\ \vspace{0.2cm} Deciphering the DNA of the heaviest quark}
\vfill
\Author{Michele Gallinaro \\{\it \small on behalf of the CMS collaboration}} 
\Address{\institute}
\vfill
\begin{Abstract}
The top quark,  the heaviest known elementary particle discovered at the Fermilab Tevatron more than twenty years ago, 
has taken a central role in the study of fundamental interactions.
Due to its large mass, the top quark provides a
unique environment for tests of the standard model. With a cumulative luminosity of more than 100~fb$^{-1}$ collected at $\sqrt{s}=7,8,13$~TeV
by each of the ATLAS and CMS experiments at the Large Hadron Collider in the first ten years of operation, 
top quark physics is probing uncharted territories in
precision and rare measurements with sensitivity to New Physics processes. 
This document summarizes the latest experimental measurements and studies of top quark properties. 
\end{Abstract}
\vfill
\begin{Presented}
$10^{th}$ International Workshop on Top Quark Physics\\
Braga, Portugal,  September 17--22, 2017
\end{Presented}
\vfill
\end{titlepage}
\def\thefootnote{\fnsymbol{footnote}}
\setcounter{footnote}{0}

\section{Introduction}

The Large Hadron Collider (LHC) is a complex and powerful tool built to explore the infinitesimally small world with the goal of understanding Nature in its complexity. The discovery of the Higgs boson in 2012, completes our current limited understanding of Nature encompassed by the standard model (SM).
Although consistent, the SM is insufficient to provide answers to fundamental remaining open questions of our puzzling Universe, where approximately 95\% of what surrounds us, still remains unexplained. 
Some of the topics spurring questions include
gravitation and the unification of forces, Dark Matter, asymmetry between matter and antimatter, mass and neutrino hierarchy. These are some of the questions that the LHC may hope to address, and top quarks may play an important role in this quest.

Top quarks are abundantly produced at the LHC in proton-proton collisions.
The top quark is the heaviest of the known fundamental particles and it was discovered at the Tevatron in 1995 by the CDF and D0 experiments~\cite{topdiscovery}.
The top quark is the partner of the bottom quark and its properties are well defined within the SM, with perhaps the notable exception of the mass which is a free parameter. The measurement of the properties of the top quark have been an intense topic of scrutiny since its discovery. Thanks to the large samples available today at the LHC, areas previously unexplored are becoming increasingly accessible. Those studies cover a large variety of measurements using basic methods, as well as increasingly sophisticated techniques to reduce the uncertainties of the measurements, and possibly maximize the sensitivity to searches of beyond the SM (BSM) processes.

Studying the top quark properties is interesting for several reasons. The top quark is almost as heavy as a gold atom, yet it behaves as a point-like particle, and it is unique among the known quarks as it decays before it can form hadronic bound states. This has important consequences,  as it offers the possibility to explore the interactions of a bare quark at energies from a few hundred GeV to several TeV. 

Studies of top quark events provide a self-consistency check of the SM, and precise measurements of top quark properties will allow to probe SM predictions.
Besides the mass, other intrinsic properties are its width, charge, and spin. Thanks to its large mass, the top quark is an excellent probe of the mechanism that breaks the electroweak gauge symmetry, and could shed light into the understanding of this phenomenon. The top quark is also a good probe for other non-SM interactions which could be induced, for example, by non-standard Higgs bosons. It could address the question whether there are there any new top quark decay modes, including supersymmetric particles, or through its branching ratios and the structure of the Wtb vertex. It is also interesting to understand whether the point-like behaviour will persist at smaller distance scales. Some of these questions can be addressed by the results obtained by the LHC experiments~\cite{atlas}.
The extent of the answers will depend on the number of events collected and on the understanding of the data themselves. 

Studies of top quark properties include those at the production, such as spin correlations, asymmetries, CP violation, or at the decay, such as branching ratios, the Cabibbo-Kobayashi-Maskawa (CKM) $V_{tb}$ matrix element, flavor-changing neutral-current, rare decays, W-boson helicity, and anomalous couplings. Any significant deviation with respect to SM expectations would be a hint of New Physics (NP).

\section{Properties at production and decay}

\subsubsection*{W boson helicity}
The W boson helicity fractions from top quark decays are measured in events with either an electron or a muon, four jets, of which two jets are identified as originating from b-quarks~\cite{whelicity}.
Top quarks decay before hadronization via $t\rightarrow Wb$, and the spin information is directly transferred to its decay products.
The strength and structure of the Wtb vertex is determined by the V-A structure. 
In the SM, W bosons are produced with left-handed ($F_L$), right-handed ($F_R$), or longitudinal ($F_0$) polarization, with the constraint that $F_0+F_L+F_R=1$. Longitudinal ($\sim 70\%$) and left-handed ($\sim 30\%$) are the dominant polarization modes. 
Experimentally, the W boson helicity can be measured through the study of the angular distribution of the top quark decay products.
The helicity angle $\theta^*$ is defined as the angle between the direction of the charged lepton from the W boson decay and the b-jet, in the rest frame of the W boson.
Charged leptons from left-handed W bosons are preferentially emitted in the opposite direction of the W boson, and thus tend to have lower momentum and be closer to the b-jet from the top quark decay, as compared to charged leptons from longitudinal or right-handed W bosons.
The predicted distributions of $\cos\theta^*$ for the different helicity fractions are compared to the data, and 
agree with the SM predictions.
The measurement is sensitive to non-SM couplings between W boson, top quark, and the bottom quark; a general parametrization of the Wtb vertex can be expressed as a function of vector and tensor couplings for left- and right-handed polarization. Using these results, limits on anomalous couplings are obtained by fixing the two vector couplings to their SM values ($V_L\equiv V_{tb}=1$, $V_R=0$), and choosing the tensor couplings as free parameters. 

\subsubsection*{$V_{tb}$ and top quark width, couplings to tau leptons and charged Higgs}
In the SM, the top quark is expected to decay as $t\rightarrow Wb$ with a branching fraction close to 100\%, as decays to a W boson and a quark 
of different isospin are strongly suppressed. 
The magnitude of the CKM matrix element $|V_{tb}|$ is expected to be close to unity as a consequence of unitarity and 
a deviation from this prediction could arise from a fourth quark generation, or simply due to different decay modes. 
The heavy flavour content of $t\bar{t}$ 
events can be measured as the ratio R=BR($t\rightarrow Wb$)/BR($t\rightarrow Wq$), where BR($t\rightarrow Wq$) is the branching 
fraction of the top quark to a W boson and a q quark (q=b, s, d). 
At the LHC, the measurement is performed by analyzing the b-tagging jet multiplicity in $t\bar{t}$ dilepton events.
Results are in good agreement with the SM prediction, 
Assuming a unitary, three-generation CKM matrix, $|V_{tb}|=1.011^{+0.018}_{-0.017}$~(stat.+syst.) is measured and $|V_{tb}|>$0.972 is obtained at 95\% CL.
The result is combined with a previous measurement of the t-channel single-top-quark cross section to determine the top-quark total decay width, $\Gamma=1.36\pm 0.02 ({\rm stat}) ^{+0.14}_{-0.11} ({\rm syst})$~GeV~\cite{vtb}.

A direct bound on the top quark decay width is also derived by partially reconstructing the kinematics of top quark candidates in dilepton final states, and comparing to simulations for different top quark width scenarios using a likelihood technique. Under SM hypotheses, 95\% limits at $0.6\le \Gamma \le 2.4$~GeV are set~\cite{topwidth}.

Couplings of the top quark to new particles can change rates, decay modes, and angular distributions of the decay products. 
As an example, the presence of a charged Higgs boson could alter the couplings and the kinematical properties (\eg polarization) of the tau resulting from the $t\rightarrow H^+ b$ decay, and stringent limits can be set on charged Higgs production~\cite{chargedhiggs1, chargedhiggs2}.

\subsubsection*{Spin correlation and polarization}
The top quark lifetime ($\sim 3\times 10^{-25}$~s) is shorter than the hadronization scale 
and it also shorter than the spin decorrelation time scale. Consequently, measurements of the angular distributions of top quark decay products give access to the spin of the top quark, allowing precise tests of perturbative QCD in top quark pair production.
For large (small) $t\bar{t}$ invariant masses, the production is dominated by the fusion of pairs of gluons with opposite (same) helicities, resulting in top quark pairs with (anti-)parallel spins in the $t\bar{t}$ center-of-mass system.

In $t\bar{t}$ production, top quarks are unpolarized but their spins are correlated. 
Measurement of the top quark pair spin correlations and top quark polarization are performed using events with two oppositely charged leptons, from the angular distributions of the two selected leptons~\cite{spincorr1}.
Measurements are performed both inclusively and differentially, for the invariant mass, rapidity, and transverse momentum of the $t\bar{t}$ system.
Spin information can be accessed through the angular distributions of the top decay products. 
A fit to the azimuthal angle difference between the leptons using a binned likelihood is performed; the measurement relies only on the leptonic information without requiring a full top quark event reconstruction.
A measurement of the top quark spin correlation is also performed in the muon+jets final state,
and the compatibility of the data with the SM hypothesis and the fully uncorrelated hypothesis are tested. The likelihood ratio for each event is calculated using a matrix element under the two hypotheses, and it is used in a template fit to extract the fraction of SM spin correlation prediction~\cite{spincorr2}.
A search for NP in the form of anomalous top quark dipole moments is performed and exclusion limits are evaluated.

Although spins are correlated, there is negligible polarization in the SM, arising from electroweak corrections to the QCD-dominated production process.
The charged lepton is the best spin analyzer among the top quark decay products, and is sensitive to the top quark spin through the helicity angle $\theta^*$.
The polarization is determined by counting the asymmetry of the number of events using the helicity angle of the positively (negatively) charged lepton in each event.

\subsubsection*{Charge asymmetry}
In $t\bar{t}$ events, the difference in rapidity (or other) distributions of top and anti-top quarks is usually known as charge asymmetry, and is sensitive to BSM models.
For example, it probes perturbative QCD predictions and provides tests of models where top quark pairs are produced through the exchange of new heavy particles, 
such as axigluons with anomalous axial-vector coupling of gluons to quarks, Z' bosons, or Kaluza-Klein excitations of gluons.

At the Tevatron, top quarks are emitted in the direction of the incoming quark, anti-top quarks in the direction of the incoming anti-quark. 
Therefore, due to the asymmetric initial state, the asymmetry manifests as a forward-backward asymmetry in the rapidity difference $\Delta y$ between top and anti-top quarks as
$A=\frac{N^+-N^-}{N^++N^-}$, where $N^+ (N^-)$ is the number of events with positive (negative) values of $\Delta y$. 
Not-so-recent Tevatron results~\cite{tevatron_asymmetry0} indicated some tension (a $3\sigma$ discrepancy) with SM predictions; 
latest results are consistent with SM predictions~\cite{tevatron_asymmetry1, tevatron_asymmetry2}.
At the LHC, in $pp$ collisions, there is no forward-backward asymmetry as the initial state is symmetric. The quantity of interest is charge asymmetry and it shows 
as a preferential production of top quarks in the forward direction due to the fact that the anti-quarks  (from the proton's sea) carry a lower momentum fraction. 
Differential measurements are obtained as a function of $p_T$, $y$, and invariant mass $M_{t\bar{t}}$ of the top quark pair. Measurements are compatible with the SM predictions~\cite{lhc_asymmetry1,lhc_asymmetry2,lhc_asymmetry3}. 
Complementarity of the results from the Tevatron and those from the LHC yield stringent limits on some BSM models\cite{lhc_asymmetry4}.

\subsubsection*{CP violation}
Violation of combined charge conjugation and parity (CP) is introduced in the SM via a phase in the CKM matrix. Detailed experimental investigation of CP violation has been conducted. Measured asymmetries are well described by the SM, but are too small to explain the observed matter-antimatter asymmetry of the Universe. However, many BSM theories predict sizable CP-violating effects.
A search in the production and decay of $t\bar{t}$ events based on the asymmetries of four observables shows no evidence for CP violating effects, consistent with SM expectations~\cite{cpviolation}.

\section{Summary}

After more than twenty years since its discovery, the top quark still plays a central role in the understanding of the SM.
Many fundamental questions still remain unanswered; 
however, some of the perhaps most interesting and still open questions can be investigated at the LHC by using top quarks as a probe of the SM.
Altered couplings, rare decays, anomalies in the properties of the top quark could all provide hints of the presence of New Physics in the data.
With the large data samples available and well understood detectors, detailed studies may shed light on important open questions.

\vspace*{-0.1cm}

\Acknowledgements
To my friends and colleagues who embarked on this long journey to decipher some of the difficult questions to which we have no easy answers. 
To Nature who keeps us wondering how all this amazing Universe remains a mystery and continuously reminds us of our limited human existence. 
To the Organizers of this series of workshops for keeping the stimulating environment in a friendly and fruitful atmosphere, where curiosity remains the main topic of discussion.

\vspace*{-0.1cm}

\begin{footnotesize}

\end{footnotesize}

\end{document}

%% file: econfmacros.tex
\def\beq{\begin{equation}}
\def\eeq#1{\label{#1}\end{equation}}
\def\eeqn{\end{equation}}

\def\beqa{\begin{eqnarray}}
\def\eeqa#1{\label{#1}\end{eqnarray}}
\def\eeqan{\end{eqnarray}}

\let\bar=\overbar

\def\eg{{\it e.g.}}

\def\Dslash{\not{\hbox{\kern-4pt $D$}}}
\def\dslash{\not{\hbox{\kern-2pt $\del$}}}

\def\msb{{\bar{\ssstyle M \kern -1pt S}}}




%% file: braga.bbl
\begin{thebibliography}{99}

\bibitem{topdiscovery}
F.~Abe {\it et al.}  [CDF Collaboration],
Phys.\ Rev.\ Lett.\  {\bf 74} (1995) 2626;\\
S.~Abachi {\it et al.}  [D0 Collaboration],
Phys.\ Rev.\ Lett.\  {\bf 74} (1995) 2632.

\bibitem{atlas}
G.~Aad {\it et al.} [ATLAS Collaboration],
JINST {\bf 3}, S08003 (2008);\\
%
S.~Chatrchyan {\it et al.} [CMS Collaboration],
JINST {\bf 3}, S08004 (2008).

\bibitem{whelicity}
V.~Khachatryan {\it et al.} [CMS Collaboration],
Phys.\ Lett.\ B {\bf 762} (2016) 512.

\bibitem{vtb}
V.~Khachatryan {\it et al.}  [CMS Collaboration],
Phys.\ Lett.\ B {\bf 736} (2014) 33.

\bibitem{topwidth}
V.~Khachatryan {\it et al.} [CMS Collaboration],
CMS-PAS-TOP-16-019.

\bibitem{chargedhiggs1}
V.~Khachatryan {\it et al.} [CMS Collaboration],
JHEP {\bf 11} (2015) 018.

\bibitem{chargedhiggs2}
V.~Khachatryan {\it et al.} [CMS Collaboration],
Phys.\ Rev.\ Lett.\  {\bf 119} (2017) 141802.

\bibitem{spincorr1}
V.~Khachatryan {\it et al.} [CMS Collaboration],
Phys.\ Rev.\ D {\bf 93} (2016) 052007.

\bibitem{spincorr2}
V.~Khachatryan {\it et al.} [CMS Collaboration],
Phys.\ Lett.\ B {\bf 758} (2016) 321.

\bibitem{tevatron_asymmetry0}
T.~Aaltonen {\it et al.}  [CDF Collaboration], 
Phys.\ Rev.\ D {\bf 83} (2011) 112003. 

\bibitem{tevatron_asymmetry1}
T.~Aaltonen {\it et al.} [CDF Collaboration],
Phys.\ Rev.\ D {\bf 87} (2013) 092002.

\bibitem{tevatron_asymmetry2}
V.~M.~Abazov {\it et al.} [D0 Collaboration], 
Phys.\ Rev.\ D {\bf 90} (2014) 072011.

\bibitem{lhc_asymmetry1}
V.~Khachatryan {\it et al.}  [CMS Collaboration],
Phys.\ Lett.\ B {\bf 760} (2016) 365.

\bibitem{lhc_asymmetry2}
V.~Khachatryan {\it et al.}  [CMS Collaboration],
Phys.\ Rev.\ D {\bf 93} (2016) 034014.

\bibitem{lhc_asymmetry3}
V.~Khachatryan {\it et al.}  [CMS Collaboration],
Phys.\ Lett.\ B {\bf 757} (2016) 154.

\bibitem{lhc_asymmetry4}
M.~Aaboud {\it et al.} [ATLAS and CMS Collaborations],
arXiv:1709.05327 [hep-ex].
  
\bibitem{cpviolation}
V.~Khachatryan {\it et al.}  [CMS Collaboration],
JHEP {\bf 03} (2017) 101.

\end{thebibliography}
